\documentclass{llncs}
	
	\usepackage[english]{babel}
    \usepackage[utf8]{inputenc}
    \usepackage[T1]{fontenc}
    \usepackage{hyperref}
    \usepackage{color}
    \usepackage{csquotes}
    \usepackage{graphicx}
    \usepackage{cite}
    \usepackage[normalem]{ulem}
    \usepackage{todonotes}
    \usepackage[misc]{ifsym}
    \usepackage{doi}

\definecolor{gray}{gray}{0.3}

\newif\ifcomments

\newcommand{\mh}[1]{%
    \ifcomments%
        \todo[color=blue!30]{\textit{Markus:} #1}%
    \fi%
}

\newcommand{\cs}[1]{%
    \ifcomments%
        \todo[color=red!30]{\textit{Christian:} #1}%
    \fi%
}

\newcommand{\notes}[1]{%
    \ifcomments%
        {\color{gray}#1}%
    \fi%
}

\begin{document}

\title{A Security Architecture for Railway Signalling}
\author{
    Christian Schlehuber\inst{1} \textsuperscript{(\Letter)} \and
    Markus Heinrich\inst{2} \and
    Tsvetoslava Vateva-Gurova\inst{2} \and\\
    Stefan Katzenbeisser\inst{2} \and
    Neeraj Suri\inst{2}
}
\institute{
    DB Netz AG, Germany\\
    \email{christian.schlehuber@deutschebahn.com} \and
    Dept of Computer Science, TU Darmstadt, Germany\\
    \email{\{heinrich, katzenbeisser\}@seceng.informatik.tu-darmstadt.de},\\
    \email{vateva@deeds.informatik.tu-darmstadt.de}, \email{suri@cs.tu-darmstadt.de}
}

\maketitle


\begin{abstract}
We present the proposed security architecture Deutsche Bahn plans to deploy to protect its trackside safety-critical signalling system against cyber-attacks.
We first present the existing reference interlocking system that is built using standard components. Next, we present a taxonomy to help model the attack vectors relevant for the railway environment. Building upon this, we present the proposed \enquote{compartmentalized} defence concept for securing the upcoming signalling systems.

\end{abstract}


\section{Introduction}
\label{sec:introduction}

The state of the art in safety-critical railway signalling typically entails the use of monolithic interlocking systems that are often proprietary, expensive and not easily exchangeable.
Consequently, the transition to more cost-effective and growth-oriented open networks is desired that can also utilize commercial off-the-shelf (COTS) hardware and software, provided the safety requirements are met.

These drivers have led Deutsche Bahn (DB) to explore transforming its signalling infrastructure using open networks and COTS to reduce cost and maintenance overhead.
At the same time, the risk of cyber-attacks introduced by open networks and COTS needs to be explicitly addressed to avoid any compromise of safety.
This work documents DB's ongoing experience in developing new signalling architectures that by-design decouple safety and security functionalities.

In this context, we first present a taxonomy of attacks outlining the potential cyber-threats relevant to protecting a railway signalling system.
Consequently, utilizing the actual layout of the currently used German railway command and control system, we propose a security architecture that explicitly delineates safety and security, and will be deployed by DB in Germany's new interlocking systems (ILS) to address security concerns.
The architecture is compartmentalized into zones and conduits following IEC~62443~\cite{iec62443}.
Thereby we regard the German prestandard DIN VDE~V~0831-104~\cite{vdev0831-104} which is a guideline to apply IEC~62443 to the railway signalling domain with respect to the very strict safety requirements.

\section{Current Interlocking Network Architecture}
\label{sec:network_model}

\begin{figure}[tb]
    \centering
    \includegraphics[width=\linewidth]{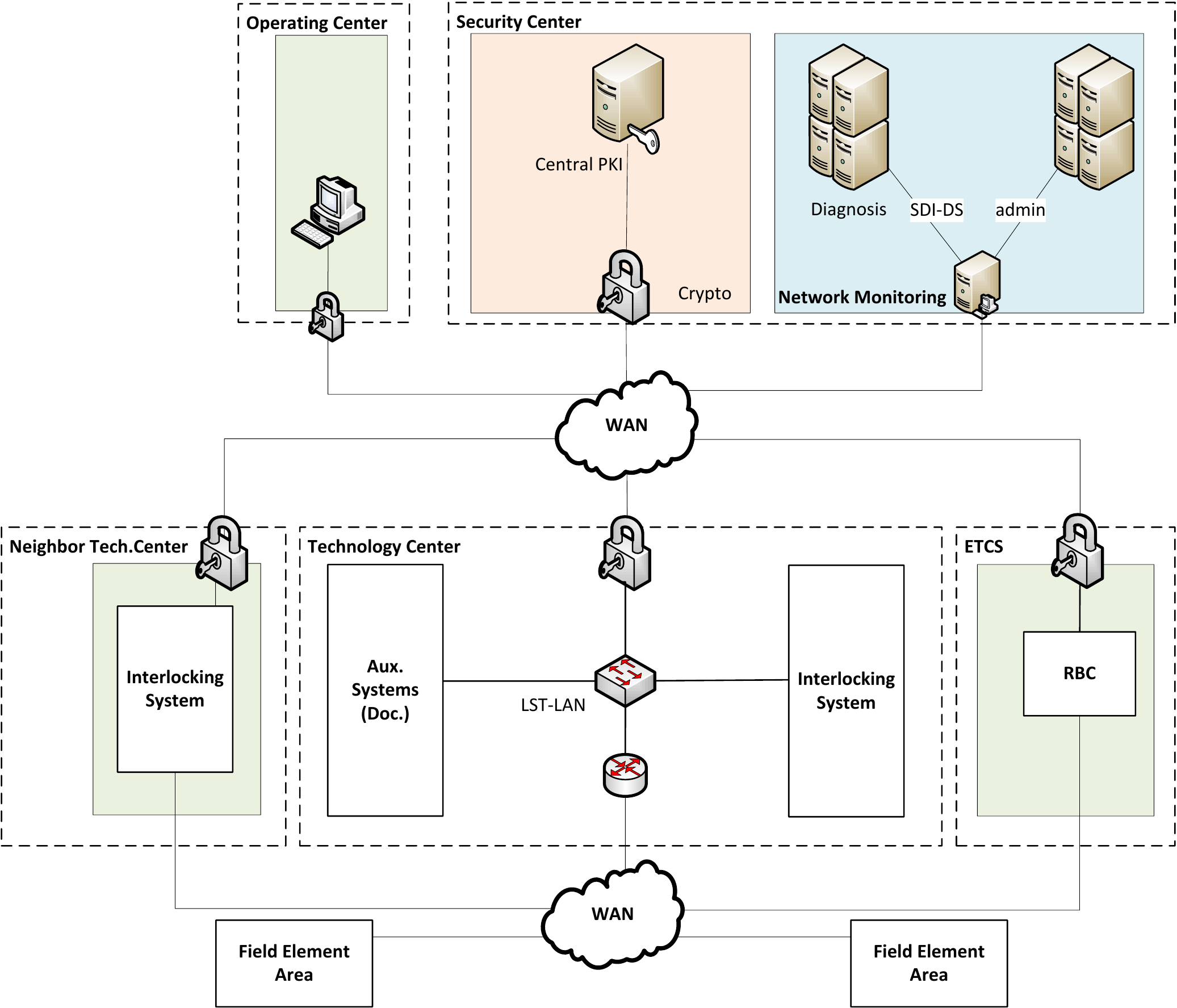}
    \caption{Architecture of a typical signalling network.}
    \label{fig:network}
\end{figure}

The reference architecture, as currently deployed by DB, is divided into three layers: \emph{Operational Layer, Interlocking Layer} and the \emph{Field Element Area}.

The \emph{Operational Layer} (upper blocks of Fig.~\ref{fig:network}) consists of an Operating Center and a Security Center\mh{Im workshop paper nennen wir es Operations Control Center, das sollten wir vereinheitlichen}.
The Operating Center is responsible for the central monitoring and controlling of the system and is equipped with central switching points.
The Security Center provides security services to the system such as security monitoring of certain communication channels and management of the Public Key Infrastructure (PKI).
As depicted in Fig.~\ref{fig:network}, the communication between the Operational Layer and the Interlocking Layer of the reference architecture is encrypted.
The Security Center has the same or higher security requirements compared to the rest of the components.

The \emph{Interlocking Layer} (middle blocks in Fig.~\ref{fig:network}) provides the safety logic of the system.
The main components of the Interlocking Layer are the Technology Center and the interface to the European Train Control System (ETCS), as depicted in Fig.~\ref{fig:network}.
The Technology Center is comprised of the ILS and auxiliary systems (e.g., needed for documenting the actions of the ILS).
The ILS plays a central role in the reference architecture by ensuring system's safety given its critical role to control signals, switches and to prevent any conflicting train movements.

The \emph{Field Element Area} (FEA) (lowest blocks in Fig.~\ref{fig:network}) provides the interface to the actual trackside signalling elements called field elements.
These are signals, points, and train detection systems amongst others that are steered by Object Controllers (OC).


Communication across the components of the Operational and Interlocking layers takes place over a Wide Area Network (WAN) through the use of Standard Communication Interfaces.
Typically, the Rail Safe Transport Application (RaSTA) Protocol~\cite{vdev0831-200} is used as a unified communication protocol for all the defined interfaces.
RaSTA targets at guaranteeing safety in the communication of railway systems.
Each RaSTA-network is assigned a network identification number which is unique within the given transport layer.
A safety code is used to guarantee the integrity of the transmitted messages.
Required redundancy for the system's high availability is omitted in Fig.~\ref{fig:network} to reduce complexity.


As can be seen in Fig.~\ref{fig:network}, only the communication between the Operational and Interlocking Layers of the reference architecture is encrypted.
This is insufficient from a security perspective, and naturally the entire communication chain across the Technology Center, the FEAs and the linking communication interfaces need to be protected.
However, enhancing the presented architecture in terms of security is not a trivial task, as various operational and compatibility constraints make introducing innovations to the interlocking system rather cumbersome.
A complicating factor being re-ensuring that no safety violations get introduced with any security related changes (i.e. proving freedom of interference).
In a normal computational environment, addressing security issues might require rapid patching and frequent updates.
However, for the safety-critical railway environment any changes to a critical infrastructure, such as the signalling system that might affect the safety of the system, require explicit approval by the National Safety Authority.
This can take significant time and exacerbates the timely reaction to security risks.
In addition, the limited hardware resources in the signalling system do not allow deploying widely-used security solutions that are computationally intensive.
Moreover, it is expected that deployed systems are used over a long operational lifetime (typically decades) and also provide strong timeliness response guarantees. All of these constraints need to be explicitly addressed when proposing a security-oriented signalling system architecture.

\section{Railway Security Assessment}
\label{sec:attack_model}

In order to propose a security architecture, this section presents the prerequisites needed for defending signalling infrastructure and also elucidates the capabilities of attackers against which the signalling system needs to be protected.
To systematically tackle the problem of enhancing security in interlocking systems, we first provide a taxonomy of the attacks relevant for the railway environment.


Given the physically large spatial scattering of the railway infrastructure, it is infeasible to install physical protection comparable to a limited area factory premise.
Access control and plant security, as important elements in a factory's security concept, do not apply to the full extent across the railway system.
Only some parts -- for example the interlocking computer -- reside in a building that offers physical perimeter protection, while others (e.g., the field elements) lie unprotected along the railtracks.



In addition, we need to ensure safety and high availability of railway signalling systems.
This is tightly coupled with the timeliness requirements of critical communication between network entities.
In cases where we cannot preclude attacks, it is necessary to install monitoring systems that can detect ongoing attacks.
For setting up a proper security concept we first need to define the capabilities of the attacker we want to defend the system against.
In the railway signalling community it is widely recognized that some security incidents are already covered by the established safety functions. 
The design of DB's security architecture follows the standard IEC~62443~\cite{iec62443} and the German prestandard DIN VDE~V~0831-104~\cite{vdev0831-104}.
They classify the strength of attackers according to their (financial) resources, their motivation, and their knowledge.
With the attacker strength in mind, we capture attacks that can be performed in a taxonomy that scopes the applicable security measures.


A taxonomy can facilitate enhancing security, as it can represent the diverse attack scenarios that threaten the railway signalling system, and also allows to consider future threats.
While sophisticated attack scenarios have been considered by the taxonomies in~\cite{hansman2005taxonomy,howard1998common,meyers2009taxonomies,simmons2009avoidit,weber1998taxonomy} as well, most of them go beyond attack vectors and include information on the targeted system~\cite{hansman2005taxonomy,simmons2009avoidit} that can be as detailed as software versions.
However, unlike contemporary taxonomies built on full information access, we consider the systems from the operator perspective and do not know beforehand which technology the vendors use to meet the requirements.
Thus, we are constrained to only model generic requirements of the systems.

\begin{figure}[tb]
    \centering
    \includegraphics[width=\linewidth]{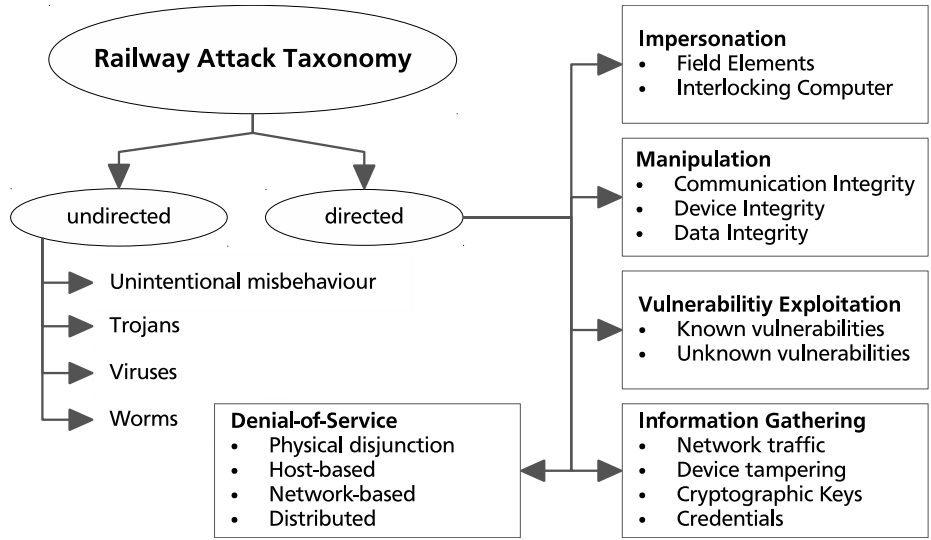}
    \caption{A Railway Attack Taxonomy.}
    \label{fig:taxonomy}
\end{figure}

Figure~\ref{fig:taxonomy} outlines our approach to categorize threats.
On the top level we distinguish across directed and undirected attacks.
This is justified by the following assumption: It is impossible for undirected attacks to cause an unsafe state in the signalling system, as they will typically not circumvent the existing safety measures.
However, this class of attacks may affect the availability of the system.

Since casualties could be the consequence, we consider impersonation as the most severe attack (i.e. an attacker being able to forge authentic messages of a network entity such as a OC or the ILS computer itself).
As in any other network that comprises standard components, all known and unknown vulnerabilities pose a threat to the system in case they are exploited.
Thus, vulnerabilities must be regarded in an attack model.
Due to the scattered physical layout of the network it is prone to many kinds of information gathering attempts, and network entities like the field elements are difficult to protect against physical tampering.
Although confidentiality is not an important target of signalling security, some information like cryptographic keys that are used to protect entities and communication channels, as well as account credentials, need to be kept secret.
A compromised key would enable more severe attacks on the system, for example impersonation.
This interconnection shows that a holistic approach is needed to secure railway signalling and neither perimeter protection nor isolated solutions will suffice.
Orthogonal to the presented threats are denial-of-service attacks where no comprehensive countermeasure exists.
The signalling systems mitigate this threat by utilizing redundancy and avoiding single points of failure.
We do not explicitly depict redundancy in Fig.~\ref{fig:network} and~\ref{fig:network_final}, though all signalling relevant communication is performed over at least two separate channels provided by RaSTA.
Entities such as the Security Center (from Fig.~\ref{fig:network}) also exist redundantly.

\section{New Security Architecture for Interlocking Systems}
\label{sec:requirements}

For safety-related railway systems, the dominant requirements are integrity, timely delivery of critical messages and system availability.
To ensure this, a Reliability, Availability, Maintainability, and Safety (RAMS) lifecycle has been introduced by EN~50126~\cite{en50126} to make the current signalling systems resilient to internal faults and human error.
However, EN~50126 does not consider attackers or malware that constitute a growing threat to all industrial control systems, including railway signalling systems.
Thus, enhanced security mechanisms are needed, provided their potential to detrimentally affect safety and availability is explicitly delineated.
This makes it infeasible to introduce standard \enquote{commercial} anti-malware and anti-virus systems into an ILS network, as the side effects are not easily discernible to be controlled.

\begin{figure}[t]
    \centering
    \includegraphics[width=\linewidth]{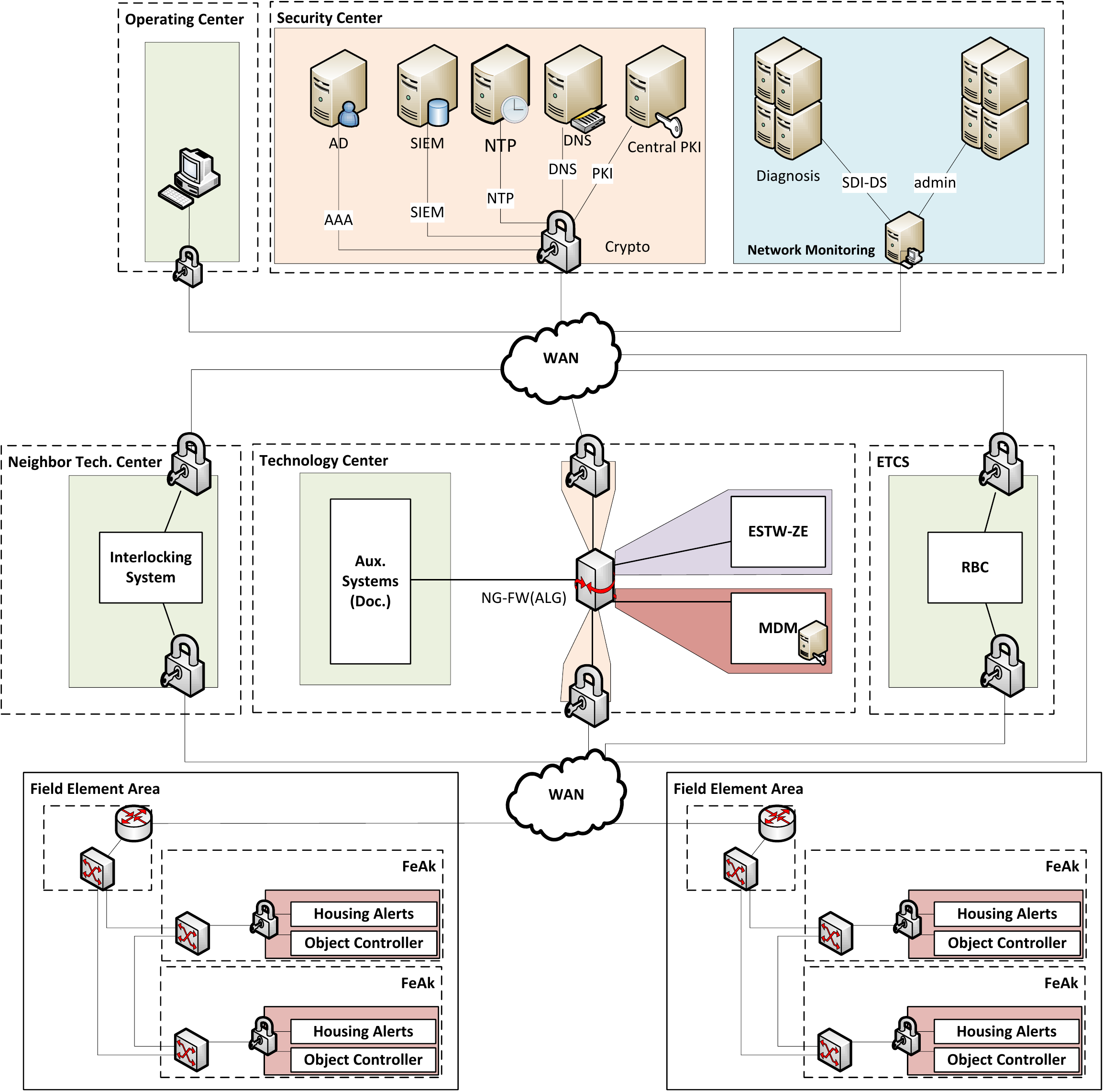}
    \caption{Proposed Security Architecture for interlocking systems of DB.}
    \label{fig:network_final}
\end{figure}

Based on the developed attack model taxonomy, a security architecture for the new interlocking technology was engineered.
The security engineering process is based on the standard IEC~62443-3-3~\cite{iec62443} with guidance taken from DIN VDE~V~0831-104~\cite{vdev0831-104}.
According to the general system design the signalling system has been partitioned into functional blocks e.g., Object Controllers (OC) and ILS (see Fig.~\ref{fig:network_final}).
The reference architecture is additionally divided into zones and conduits, where each zone is logically or physically defined~\cite{iec62443}.
According to IEC~62443 each object within the architecture being hardware, software, user, etc. is assigned to exactly one zone or to exactly one conduit.
A zone (colored areas) is a grouping of assets that have common security requirements which is expressed as a Security Level (SL) that is assigned to each zone.
Conduits are the communication channels between zones with both the same and different security requirements.

A risk analysis yielded SLs of 2 or 3 for every zone.
Based on these SLs, the security requirements were defined for every component of the system to ensure the fulfilment of a defence-in-depth concept.
The requirements range from password changing abilities over cryptographic functions to a set of requirements that support the later detection and analysis of attacks e.g., logging capabilities.

After the zones have been provided with security measures, the conduits between them remain a vulnerable point.
In contrast to the zones, \mbox{IEC~62443-3-3} does not contain guidance on how to secure conduits.
Over our requirements and taxonomy process two types of conduits have been identified, namely: (a) conduits connecting zones of equal SL, and (b) conduits connecting zones of different SLs.
Conduits which only have unidirectional data flow could also be considered, but these are only a subtype of one of the former described conduits.

The system layout of Fig.~\ref{fig:network} has been extended to secure the zones and conduits, as shown in Fig.~\ref{fig:network_final}.
Again, redundancy is omitted.
The FEA is provided in more detail to show the security application.
Multiple OCs are presented as there are a number of field elements to steer in a single FEA.
For redundancy, they are organized in a ring topology with switches (angular boxes) and routers (round boxes).
The relation between OC and field element is usually one-to-one.
Security boxes have been added to every OC (depicted as locks) in the FEA within a junction box (labelled FeAk).
They provide the system with encryption capabilities and the possibility for basic filtering and DoS prevention rules.
The capabilities are required for securing conduits between zones with equal SLs.
The boxes are based on a ruggedized and hardened hardware platform.
As they are completely separated from the safety functionality, they can be applied as a replacement of switch components in the interlocking network and even be introduced during system upgrades.
The security terminates in the security box, thus the safety hardware need to be protected by physical measures.
The FEA junction boxes are thus physically protected by \enquote{housing alerts} that trigger an alarm to prohibit attackers from tampering with the system.

In the Technology Center, a termination point for the field element encryption has been introduced.
Also, several zones with different SLs have to be connected, e.g., the interlocking system has to be connected to the maintenance and data management subsystem (MDM) with different SL.

To tackle this challenge, an application layer gateway (ALG) has been introduced as a central entity of the Technology Center.
This device is configured to only allow desired connections between zones.
Via packet inspection mechanisms malicious code can be identified.
If zones of different SL are connected, the allowed communication can be limited via white-list filtering on different layers.
If anomalous behaviour is detected the ALG reports this to the Security Operation Center (SOC), where an operator can decide what actions have to be taken.
In certain cases the separation of a zone from the rest of the network (quarantine) may be needed, which then can be realized by the ALG.
Upon the detection of new attack scenarios the operator also has the possibility to change the rule set and filtering of the ALG to mitigate the new attack.

On the operational layer the SOC has been extended by a Security and Information Event Management (SIEM) system besides elements for system management, such as PKI, domain name service, network time server, and a directory service.
The SIEM system aggregates information from every component and analyses it for possible attacks.
If it detects a possible attack the security operator is informed, starts with further investigation on the issue, and finally performs some action to solve it.

As the provisioning of security requires the application of tools and methods on a sustained basis, a process based approach is implemented to ensure a constant level of security.
For this a patch management process has been developed.
Changes to components are first checked in a simulated environment for quality assurance before they are applied to the operational components.
For a rapid reaction to attacks, the rule sets of the ALG and security boxes can be altered to mitigate the vulnerability until a patch can be applied.
Furthermore, processes for incident management and an Information Security Management System (ISMS) have been implemented.
Upon the detection of an anomaly it is checked against a database of known incidents and relevant actions are applied.
For unclassified anomalies, forensics are performed to determine the relevant reaction.
After solving the incident, the findings are used as input for the ISMS to enhance the security processes.

By having added security features to the communication channel of the safety building blocks, the architecture allows to control that strict safety requirements such as availability and timeliness are still met.
The communication channel is transparent to the safety system such that the security blocks can be updated independently and without affecting the safety homologation process.
The decoupling of safety and security still requires to make the physical gap between them as small as possible (e.g., on the same circuit board), to avoid attacks just behind the security component.

\cs{
Should we also elaborate on the processes or only say, that they exist?
}
\mh{No, we do not have much space. What is already there should be enough}

\notes{
\begin{itemize}
\item connection to SIEM and CSIRT
\item long term patches; fast conf updates of FW
\item recovery processes (?)
\end{itemize}
}


\section{Conclusion}
\label{sec:conclusion}
The existing interlocking architecture provides insufficient security against cyber-attacks.
To overcome this, DB plans to deploy the presented security architecture in Germany's new ILS to mitigate security risks without detrimentally impacting the system's safety.
The presented security concept includes monitoring and information systems as well as basic security building blocks such as cryptography support and filtering.
It ensures security not only cross the Operational and Interlocking layers but also provides security functions for the Technology Center and the Field Element Areas.
In addition, processes are established to ensure the correct handling of incidents and functional requirements to each building block in order to help build security enabled components.

\subsubsection{Acknowledgements.}
\label{ssub:acknowledgements}
Research supported in part by EC CIPSEC GA 700378.

The final publication is available at Springer via \url{http://dx.doi.org/10.1007/978-3-319-66266-4_21}.

\bibliographystyle{splncs}
\bibliography{bibliography}

\end{document}